\documentclass[final,12pt,3p]{elsarticle}

\graphicspath{{./Images/}} 

\usepackage{amssymb}

\usepackage{color}

\usepackage[squaren, Gray, cdot]{SIunits}

\begin{document}

\begin{frontmatter}



\title{Uniaxial anisotropy and enhanced magnetostriction of CoFe$_2$O$_4$ induced by reaction under uniaxial pressure with SPS}


\author{A. Aubert*}

\cortext[cor1]{Corresponding author\\A. Aubert et al. / Journal of the European Ceramic Society 37 (2017) 3101-–3105\\ DOI : http://dx.doi.org/10.1016/j.jeurceramsoc.2017.03.036}
\ead{alex.aubert@satie.ens-cachan.fr}

\author{V. Loyau, F. Mazaleyrat and M. LoBue}

\address{SATIE UMR 8029 CNRS, ENS Cachan, Universit\'{e} Paris-Saclay,\\ 61, avenue du pr\'{e}sident Wilson, 94235 Cachan Cedex, France}

\begin{abstract}
In this study, we have compared magnetic and magnetostrictive properties of polycrystalline CoFe$_2$O$_4$ pellets, produced by three different methods, focusing on the use of Spark Plasma Sintering (SPS). This technique allows a very short heat treatment stage while a uniaxial pressure is applied. SPS was utilized to sinter cobalt ferrite but also to make the reaction and the sintering (reactive sintering) of the same ceramic composition. Magnetic and magnetostrictive measurements show that the reactive sintering with SPS induces a uniaxial anisotropy, while it is not the case with a simple sintering process. The induced anisotropy is then expected to be a consequence of the reaction under uniaxial pressure. This anisotropy enhanced the magnetostrictive properties of the sample, where a maximum longitudinal magnetostriction of $-229$~ppm is obtained. This process can be a promising alternative to the magnetic-annealing because of the short processing time required (22 minutes).

\end{abstract}


\end{frontmatter}



\section{Introduction}

In the recent years, there has been an increasing interest in improving magnetostriction of oxide-based materials, which are suitable alternative for rare earth alloys (such as Terfenol-D) due to their low cost, ease of fabrication and high electrical resistivity. Polycrystalline cobalt ferrite is an excellent candidate because various techniques of preparation permits an enhancement of the maximum longitudinal magnetostriction and piezomagnetic coefficient (d$\lambda$/d$H$). Both properties are essential to obtain actuators and sensors exhibiting great performances, which are the main applications for these materials. To achieve high magnetostrictive properties, the most common technique is to induce a magnetic anisotropy by applying a strong magnetic field during an annealing between 300 and 400~$^\circ$C~\cite{lo2005, zheng2011, khaja2012, muhammad2012, khaja2014}. This permits a rearrangements of Co and Fe ions~\cite{na1994, slonczewski1958} and leads to a uniaxial anisotropy parallel to the direction of the magnetic annealing field, hence tuning the magnetostrictive properties. Wang \textit{et al}.~\cite{wang2016} proposed another technique in which particles were oriented through a magnetic field before the sintering, thus introducing a texture in the polycrystalline sample, which also contributes to better magnetostrictive properties. In this work, a new technique that induces uniaxial anisotropy is reported, based on a reaction under uniaxial pressure using Spark Plasma Sintering (SPS) method of production. SPS process allows the fabrication of high-density bodies at much lower temperature with short processing time. During the procedure, a high uniaxial pressure is applied while a pulsed electric current heats up the die and the sample~\cite{munir2006}. SPS can be used either to activate the reaction~\cite{orru2009} or to sinter~\cite{millot2007, gaudisson2014} oxide-based materials. This paper will focus on the effect of reaction and/or sintering of the cobalt ferrite by SPS. Magnetic and magnetostrictive behavior of the distinct samples are then compared regarding the process of fabrication employed.


\section{Experimental Details}
\subsection{Samples Fabrication}
\begin{figure}[t]
\centering
\includegraphics[angle=-90,scale=0.4]{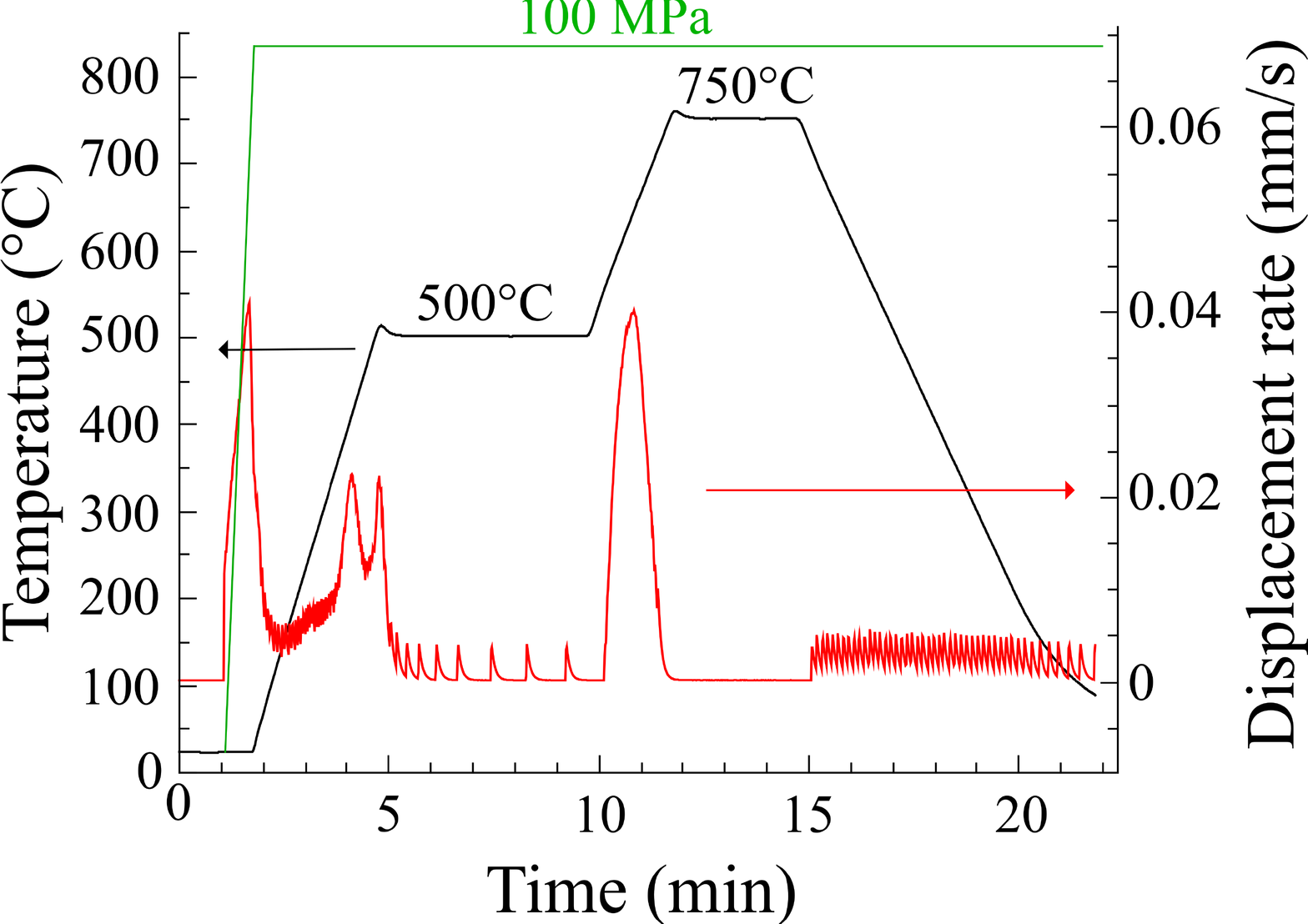}
\caption{Temperature (black), Pressure (green) and Displacement rate (red) profiles for the SPS process of the CF-RS-SPS sample. Stage at 500~$^\circ$C correspond to the reaction and stage at 750~$^\circ$C to the sintering.}
\label{ProcessSPS}
\end{figure}

Polycrystalline CoFe$_2$O$_4$ samples were prepared by three different methods. In all the cases, nanosize ($<$ 50~nm) oxides Fe$_2$O$_3$ and Co$_3$O$_4$ (Sigma-Aldrich) were used as precursors in  molar ratio of 3:1. Powders were mixed in a planetary ball mill during 30~min at 400~rpm, and then grinded during 1~hour at 600~rpm. Initially, the classic ceramic method was used to produce our sample. Mixture was first calcined at 900~$^\circ$C during 12~hours to form the spinel phase, and then grinded at 550~rpm during 1~hour. After uniaxial compaction at 50~MPa in a cylindrical die of 10~mm diameter, sample was sintered at 1250~$^\circ$C during 10~hours. This sample will be referred as CF-CM.
In the second method, the synthesis of the spinel phase was achieved under the same condition as for the ceramic method. However, the sintering process was done by SPS. In all SPS experiments, a graphite die of 10~mm diameter was used and the heating was carried out under neutral atmosphere (argon). The sintering was performed under a pressure of 100~MPa, with a 5 minutes temperature ramp from 20~$^\circ$C to 980~$^\circ$C followed by a stage of 2 min at 980~$^\circ$C before cooling down. This sample will be referred as CF-S-SPS.
Finally, in the last method, the SPS was utilized to make both the synthesis and the sintering (reactive sintering). The reaction stage was performed at 500~$^\circ$C for 5~min and the sintering stage at 750~$^\circ$C for 3~min, both under a pressure of 100~MPa. The thermal cycle was chosen based on the observation of the displacement rate of the pistons versus the temperature, as shown in Fig.~\ref{ProcessSPS}. We assume that when the displacement rate brings back down, this signify that the reaction or sintering stage is well advanced, meaning that the temperature is properly chosen. This sample will be referred as CF-RS-SPS. Regardless of the method used, cylindrical pellets of 10~mm diameter and 2~mm thick were obtained. 

\subsection{Measurement Procedures}
The crystal structures of the ceramics were characterized by X-Ray Diffraction (XRD). XRD patterns of the samples were purchased from the pellets' surfaces and the experimental instrument employed is a Bruker D2 phaser 2nd Gen diffractometer using CoK$\alpha$ radiation. Diffraction patterns were recorded in the angular range from 15$^\circ$ to 100$^\circ$ with a scan step size of 0.02$^\circ$. The refinement is done by applying the Rietvield Method using MAUD software. 

The surface morphology are analysed using scanning electron microscope (SEM)
Hitachi S-3400N model.
A hydrostatic balance was utilized to determine the density of our ceramics. The magnetic measurements were carried out on samples, cut into cube shape of 8~mm$^3$, using a vibrating sample magnetometer (VSM, Lakeshore 7400) up to a maximum field of 1~T. Magnetostriction measurements were performed at room temperature by the strain gauge method with an electromagnet supplying a maximum field of 700~kA/m. The gauges were bonded on the pellets' surface along the direction (1) and the magnetic field was applied in the three directions (1), (2) and (3) of the Cartesian coordinate system, (1) and (2) being in the plane of the disc and (3) out of plane.

\section{Results and discussion}
\subsection{Microstructure}
\begin{figure}[htb!]
\centering
\includegraphics[angle=180,width=0.48\textwidth]{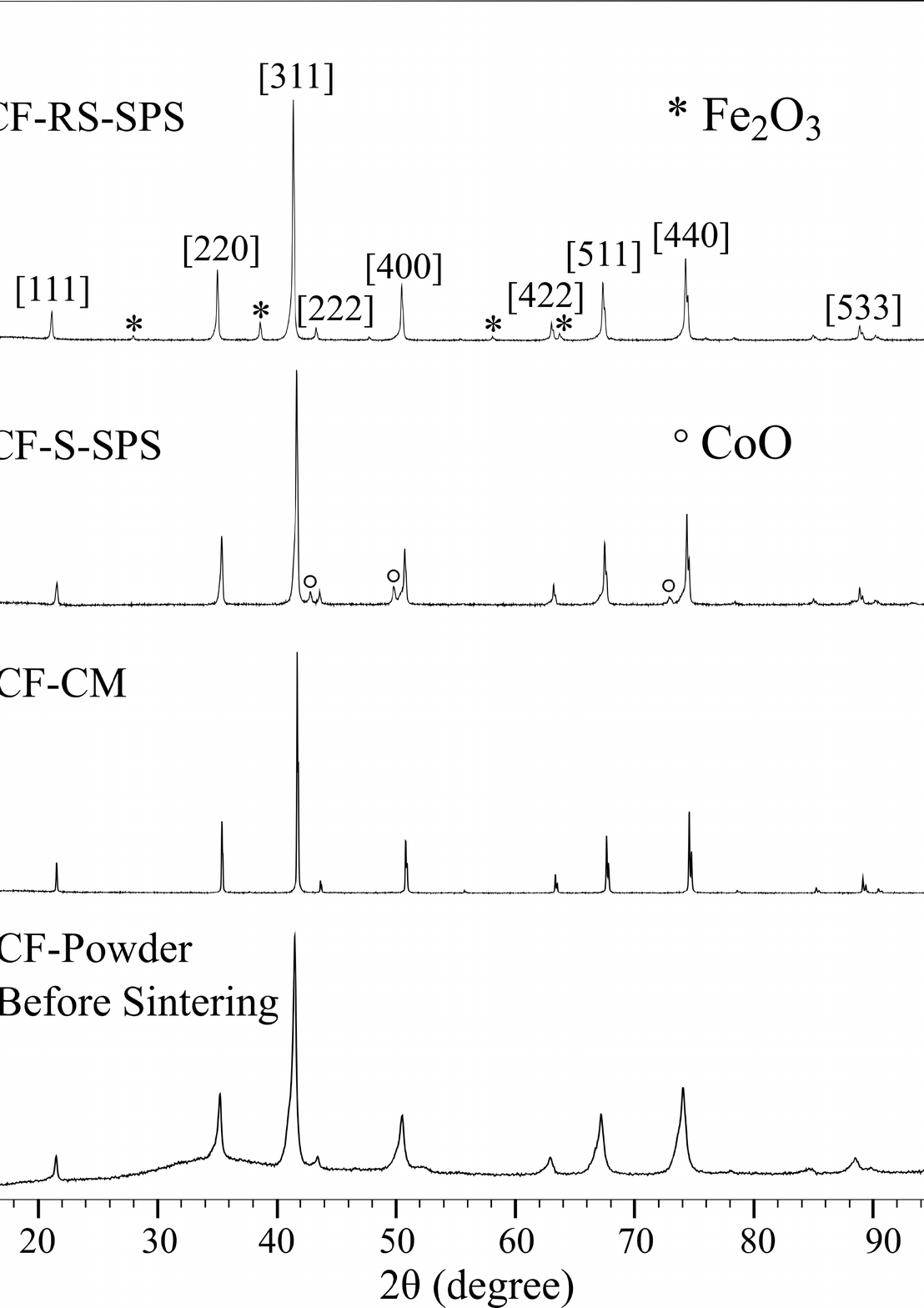}
\caption{XRD patterns of CoFe$_2$O$_4$ samples CF-RS-SPS, CF-S-SPS and CF-CM. 
The XRD results of the cobalt ferrite powder after calcination is also plotted.
}
\label{DRX}
\end{figure}

All samples were initially characterized by X-Ray Diffraction analysis, and in all cases the desired cobalt ferrite spinel structure has been obtained, as shown in Fig.~\ref{DRX}.

To make sure that no secondary phase was present after the calcination and before the sintering process of CF-CM and CF-S-SPS, the XRD analysis was also performed on the cobalt ferrite powder (Fig.~\ref{DRX}) and pure spinel phase was obtained. However, on the ceramics, the only sample free from secondary phase is CF-CM. Purity of the spinel phase for each sample are reported Table~\ref{Table1}.
CF-S-SPS sample sintered at 980~$^\circ$C presents a small amount of CoO phase (7~wt\%). This secondary phase, already reported in previous papers~\cite{millot2007,gaudisson2014}, might be a result of partial reduction of the ferrite to CoO in the graphite die during the SPS sintering. 
CF-RS-SPS sample, obtained by reactive sintering, shows 9~wt\% of hematite (Fe$_2$O$_3$). This can be a consequence of the precursor oxides Fe$_2$O$_3$ that did not completely react with the Co$_3$O$_4$ during the short reaction stage (5 min). 

\begin{figure*}[h]
\centering
\includegraphics[angle=-90,width=1\textwidth]{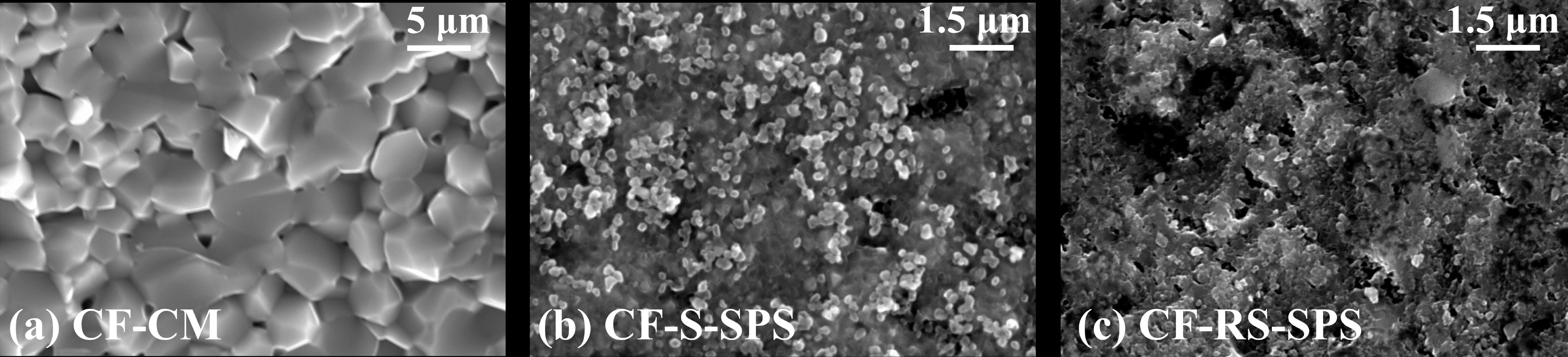}
\caption{
SEM images of samples (a) CF-CM, (b) CF-S-SPS and (c) CF-RS-SPS.
}
\label{SEM}
\end{figure*}

From MAUD refinement, it was possible to retrieve the average crystallite size (precisely the size of coherent diffraction domain $<L_{XRD}>$) for each sample (see in Table~\ref{Table1}). As expected, the size decreases with the reaction time and sintering time of the sample. To investigate the microstructure of the produced materials in more details, SEM observations were performed. The recorded SEM micrographs for the three different samples CF-CM, CF-S-SPS and CF-RS-SPS are shown in Fig.~\ref{SEM}. It is apparent that CF-CM's grain size are much larger than for the sample CF-S-SPS, which was sintered with SPS. On the other hand, the grain size of CF-RS-SPS is not easily visible since SPS permits a short reaction time (5 min) thus little grain growth~\cite{orru2009}, the grain size might be of the order of 50~nm, as the precursor oxides, which is too small to be seen with our SEM model. This goes along with the crystallite size estimated previously for CF-RS-SPS of 100~nm. Hence, only the grain size of CF-CM and CF-S-SPS are reported Table~\ref{Table1}.
Density of the ceramics were also measured and it appears that sintering with SPS techniques permits higher density (97~\%) than with the ceramic method (90~\%). It is worth noticing that XRD patterns show no significant difference in the relative intensities of the peaks for the samples. This similarity indicates that, apparently, no texture was induced during reactive sintering (CF-RS-SPS). In fact, texturing would improve the peak intensity of specific crystallographic families, which is not the case here.

\begin{table*}[ht]
\caption{Results of structural and magnetic measurements of CF-CM, CF-S-SPS and CF-RS-SPS. Properties reported are : purity of the phase CoFe$_2$O$_4$, crystallite size ($<L_{XRD}>$), grain size ($<D_{SEM}>$),  relative density (RD) of the sample, coercive field ($H_c$), remanent magnetization along the hard axis ($M^{HA}_r$) and the easy axis ($M^{EA}_r$), and saturation magnetization ($M_s$).}
\centering
\begin{tabular}{|c|c|c|c|c|c|c|c|c|}
  \hline\hline
  & Purity & $<L_{XRD}>$ & $<D_{SEM}>$ & RD & $H_c$ & $M^{HA}_r$ & $M^{EA}_r$ & $M_s$\\
  & (wt\%) & (nm) & (\micro\meter) & (\%) & (kA/m) & (mT) & (mT) & (mT)\\
  \hline
  CF-CM & 100 & 250 $\pm$~25 & 4.2 $\pm$~0.4 & 90 & 21 & 102 & 102 & 510\\
  CF-S-SPS & 93 & 120 $\pm$~12 & 0.3 $\pm$~0.1 &97 & 19 & 229 & 229 & 505\\
  CF-RS-SPS & 91 & 100 $\pm$~10 & & 97 & 53 & 205 & 301 & 452\\
  \hline\hline
\end{tabular}
\label{Table1}
\end{table*}

\subsection{Magnetism}
\begin{figure*}[htb!]
\centering
\includegraphics[width=1\textwidth]{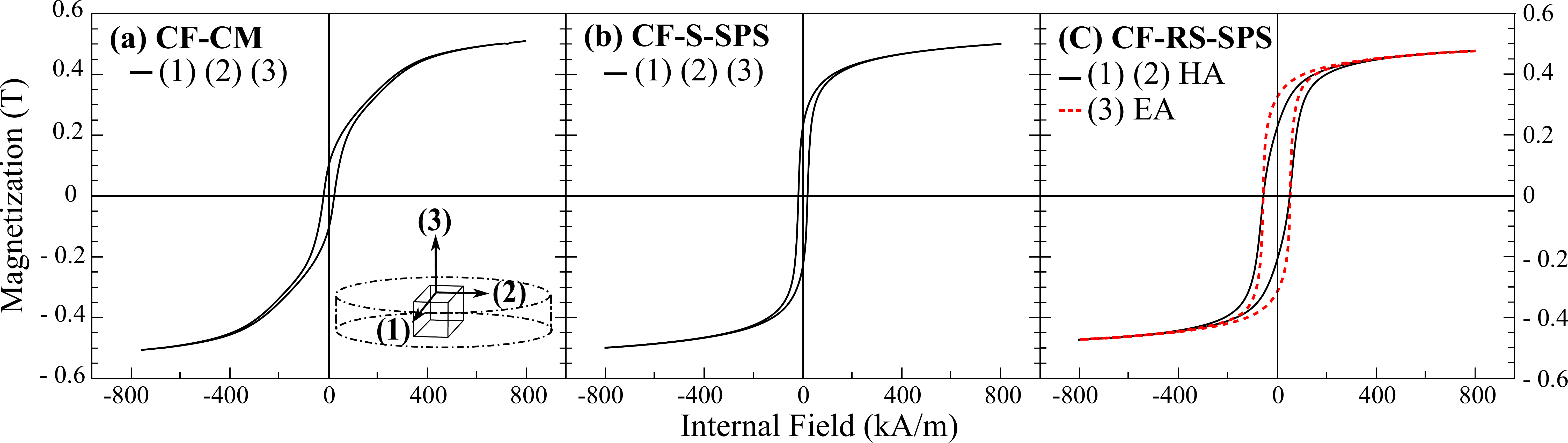}
\caption{Hysteresis loop M-H of samples (a) CF-CM, (b) CF-S-SPS and (c) CF-RS-SPS cut into cube. Measurements are done in the three directions of the cube (1), (2) and (3) as represented on the drawing.}
\label{VSM}
\end{figure*}

The magnetic hysteresis loops of the three samples are shown in Fig.~\ref{VSM}. Measurements were performed on cubes because this shape exhibits the same demagnetizing factor in the three directions~\cite{chen2005}. Thus, the magnetometric demagnetizing coefficient of a cube  ($N_m = 0.2759$) was taken into account to plot the curves as a function of the internal field. The magnetic measurements were done in the three directions (1), (2) and (3) of the cube as sketched in Fig.~\ref{VSM}~(a). CF-CM and CF-S-SPS are represented in Fig.~\ref{VSM}~(a) and Fig.~\ref{VSM}~(b), respectively. As they exhibit considerably similar loops in the three directions with the same remanent magnetization ($M_r$) and coercive field ($H_c$), we have represented only one curve out of three. This shows the isotropic behavior of such ceramics. On the other hand, it is apparent that the M-H loops of CF-RS-SPS in Fig.~\ref{VSM}~(c) present uniaxial magnetic anisotropy in the direction (3), because the remanent magnetization is higher than that for the directions (1) and (2). 

The values of coercive field ($H_c$), remanent magnetization ($M_r$) for the easy/hard axis, and saturation magnetization ($M_s$) are reported in Table~\ref{Table1}. 
The difference in coercive field ($H_c$) and saturation magnetization ($M_s$) for CF-RS-SPS compared to the other two ceramics is most likely due to the the secondary phase Fe$_2$O$_3$. Indeed, it has been reported that the impurity Fe$_2$O$_3$ has a direct impact on the magnetic properties of the cobalt ferrite by increasing its coercive field and decreasing the saturation magnetization~\cite{nlebedim2012Co}. Here, the change in coercive field could also be a consequence of the discrepancy of the grain size. On the other hand, CF-S-SPS has similar values of coercive field and saturation magnetization compared with CF-CM despite the presence of CoO. This is because a small amount of CoO (7 wt\%) has a very low impact on these two magnetic properties~\cite{nlebedim2013Co2}. One can notice that the susceptibility at low field is higher for CF-S-SPS ($\chi=20$) than for CF-CM ($\chi=4.8$). This could be a consequence of the higher density of the sample sintered with SPS and it lower grain size~\cite{nlebedim2014}. Also, CF-RS-SPS exhibit very high susceptibility in the easy axis direction ($\chi=26.3$), but lower suceptibility in the hard direction ($\chi=6.2$). These results from the uniaxial anisotropy found for this sample.
It is interesting to note that the magnetic anisotropy appears when the synthesis of the spinel phase is performed with SPS, and does not occur when there is only a sintering stage with SPS. Thus, the key step that induces the magnetic anisotropy is the reaction during the SPS process. 

It is known that anisotropy and magnetostriction of cobalt ferrite stem from the spin-orbit coupling of Co$^{2+}$ ions, usually distributed randomly among  the octahedral sites (B sites). Some authors~\cite{zheng2011,na1994, slonczewski1958, ding1995} reported that when a magnetic annealing is done on CoFe$_2$O$_4$, the superimposed induced uniaxial anisotropy is a consequence of Co cations diffusion to particular B sites, thus leading to a preferential magnetic axis close to the magnetic field applied during the annealing.
In our case, the magnetoelastic coupling is involved by the uniaxial pressure applied on the magnetostrictive material, thus promoting a preferred orientation of the magnetic moments~\cite{dutremolet1993}. Moreover, the temperature of the synthesis stage with SPS is 500~$^\circ$C (see in Fig.~\ref{ProcessSPS}), keeping the material below the theoretical Curie temperature of the cobalt ferrite (520~$^\circ$C).
In this way, when the reaction stage is performed under uniaxial pressure, the material is ferrimagnetic and the applied stress influences the position of Co$^{2+}$ ions leading to a uniaxial anisotropy in the direction of the pressure. This could explain why the direction (3) is the easy axis in the M-H loop of the CF-RS-SPS. It also justifies why there is no induced anisotropy when SPS sintering is performed on the already formed phase. In fact, during the classical reaction of the sample CF-S-SPS, Co$^{2+}$ ions migrate randomly to the different B sites  in an equilibrium position and they remain pinned there during the sintering because the stage is too short (2~min) to permit the diffusion.

\subsection{Magnetostriction}

\begin{figure*}[!t]
\centering
\includegraphics[width=1\textwidth]{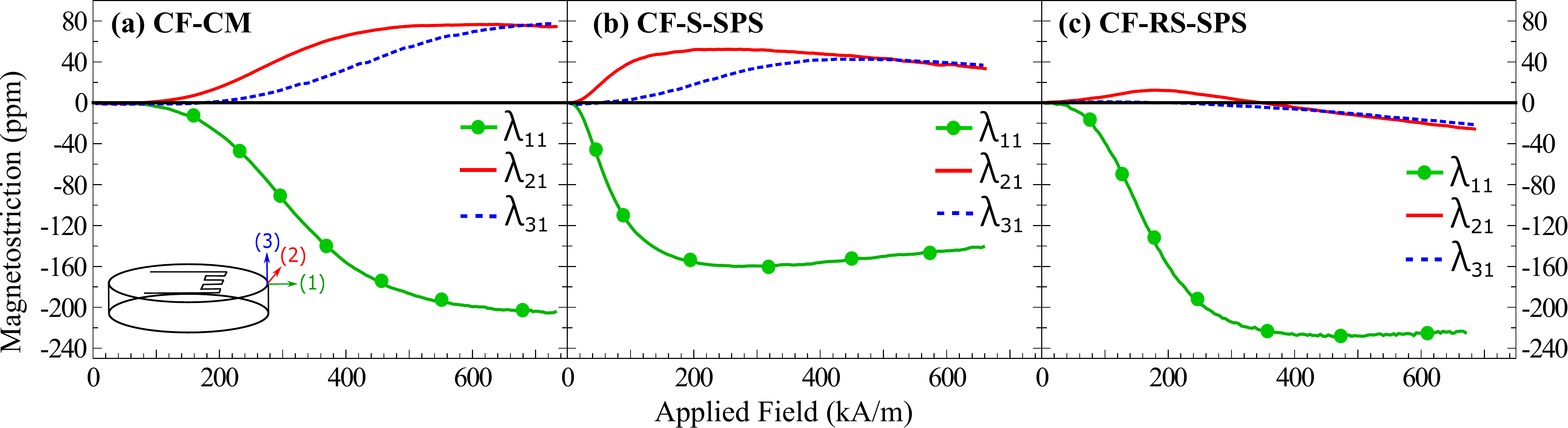}
\caption{Magnetostriction curves of (a) CF-CM, (b) CF-S-SPS and (c) CF-RS-SPS. The green line with circles ($\lambda_{11}$) correspond to the measurement when the applied field is along the direction (1), the red solid line ($\lambda_{21}$) when the applied field is along the direction (2) and the blue dotted line ($\lambda_{31}$) when the applied field is along the direction (3). The strain gauge is bonded along the direction (1) for all measurements.}
\label{FigMagnetostriction}
\end{figure*}

The results of magnetostriction measurements at room temperature are shown in Fig.~\ref{FigMagnetostriction}. Magnetostriction is always measured along the direction (1) but the magnetic field is applied in the three directions (1), (2) and (3), thus leading to $\lambda_{11}$, $\lambda_{21}$ and $\lambda_{31}$ respectively. We observe negative $\lambda_{11}$ and positive $\lambda_{21}$ and $\lambda_{31}$ in the case of CF-CM and CF-S-SPS. The maximum longitudinal magnetostriction for CF-CM is $-204$~ppm and for CF-S-SPS is $-161$~ppm. 

The reduction of the saturation magnetostriction for CF-S-SPS is possibly a consequence of the secondary phase CoO. In fact, this phase was found to affect strongly the saturation magnetostriction~\cite{nlebedim2013Co2}. Moreover, density does not seem to be the determining factor for the amplitude of the magnetostriction in our study, as it was also reported in other studies~\cite{nlebedim2014, nlebedim2010process, khaja2013}.
However, the ratio between longitudinal and transverse magnetostriction is of approximatively 3:1 for both samples. The change in slopes present at high fields are due to the contribution of the positive magnetostriction constant $\lambda_{111}$ of the cobalt ferrite~\cite{muhammad2012}. These data confirm the isotropic behavior of such ceramics and corroborate the magnetic hysteresis loops presented in the previous paragraph.
The magnetostriction curves of CF-RS-SPS are quite different compared to the other two samples, especially $\lambda_{21}$ and $\lambda_{31}$. The maximum longitudinal magnetostriction is enhanced to $-229$~ppm while $\lambda_{21}$ and $\lambda_{31}$ become negative once a certain magnetic field is reached. The ratio between longitudinal and transverse magnetostriction is of 19:1, which is much more than the theoretical isotropic value of 2:1. This type of curves is characteristic of cobalt ferrite with an induced uniaxial anisotropy along the direction (3) and it has been reported in several papers in the case of CoFe$_2$O$_4$ after magnetic annealing~\cite{khaja2012, muhammad2012}. The change in sign of both $\lambda_{21}$ and $\lambda_{31}$ might be a consequence of the modification in contribution of the two parameters that define polycristalline magnetostriction, mainly $\lambda_{100}$ at low field and $\lambda_{111}$ at high field. The magnetostrictive curves of CF-RS-SPS also agree with the magnetic measurements of such sample, where an induced uniaxial anisotropy along the direction (3) was observed.

\begin{figure}[!t]
\centering
\includegraphics[width=0.5\textwidth]{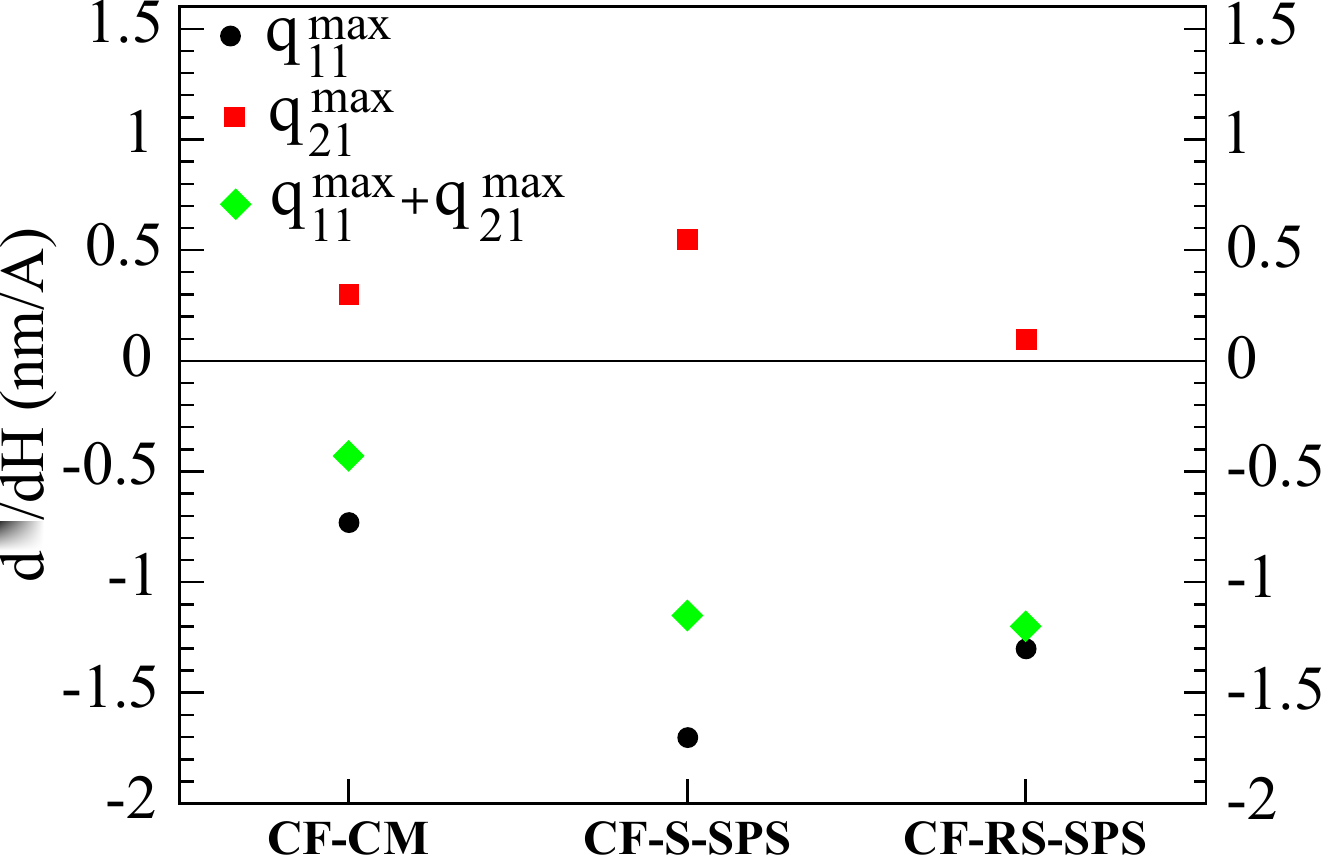}
\caption{
Maximum piezomagnetic coefficient obtained in the longitudinal ($q_{11}^{max}$) and transverse ($q_{21}^{max}$) direction for CF-CM, CF-S-SPS and CF-RS-SPS. The sum of both coefficients $q_{11}^{max}+q_{21}^{max}$ is also plotted.
}
\label{FigPiezomag}
\end{figure}

Another figure of merit for magnetostrictive materials is the piezomagnetic coefficient (or strain derivative) defined as the slope of the magnetostrictive coefficient $q^m=d\lambda/dH$. In magnetoelectric layered devices, the transverse ME effect depends directly on the sum of the longitudinal and transversal piezomagnetic coefficient $q_{11}^{m}+q_{21}^{m}$~\cite{loyau2015a}. In Fig.~\ref{FigPiezomag}, the maximum piezomagnetic coefficient $q_{11}^{max}$, $q_{21}^{max}$ and the sum of both $q_{11}^{max}+q_{21}^{max}$ are represented for the three samples CF-CM, CF-S-SPS and CF-RS-SPS. It is interesting to note that CF-S-SPS has the highest $q_{11}^{max}$ (-1.7~nm/A) compared to CF-CM (-0.73~nm/A) and CF-RS-SPS (-1.3~nm/A). This high longitudinal strain derivative for CF-S-SPS might be a consequence of its high permeability at low applied field. The magnetostriction being a quadratic function of the magnetization,~\cite{dutremolet1993, jiles1995, fetisov2015} the slope is hence directly influenced by the permeability. But as CF-S-SPS is isotropic, it also exhibits the highest $q_{21}^{max}$ (0.55~nm/A). On the other hand, CF-RS-SPS is anisotropic, resulting in a very low $q_{21}^{max}$ (0.1~nm/A). Hence, by summing up $q_{11}^{max}$ and $q_{21}^{max}$, it results in a $q_{11}^{max}+q_{21}^{max}$ lower for CF-S-SPS (-1.15~nm/A) than for CF-RS-SPS (-1.2~nm/A). For magnetoelectric purpose, CF-RS-SPS is then expected to exhibit a better effect than CF-CM or CF-S-SPS.

\section{Conclusion}

In summary, we compared the magnetic and magnetostrictive properties of cobalt ferrite discs obtained with three different methods. It has been demonstrated that samples made by the classic ceramic method and the classic reaction plus sintering with SPS behave as near-isotropic materials, while the reactive sintering with SPS induced a uniaxial anisotropy. An easy direction is found in the M-H loops, parallel to the applied pressure during the synthesis with SPS. This has a direct effect on the magnetostrictive behavior, where measurements in the three directions gave magnetostriction of the same sign once a high magnetic field is applied. It also enhances the maximum longitudinal magnetostriction of the sample. 

This enhancement could be even improved by optimizing the SPS processing parameters and hence trying to reduce the secondary phase Fe$_2$O$_3$ or increasing the uniaxial anisotropy.
Anyway, ceramic with such properties could be of great interest for magnetoelectric composites to improve their performances. The reactive sintering at SPS can be a promising alternative to the magnetic annealing process because the time required to produce a sample is much shorter than any other technique (22 min for the reaction and the sintering).

\newpage
\section*{References}

\bibliographystyle{elsarticle-num} 

\end{document}